\documentclass[twoside]{article}
\usepackage[utf8]{inputenc}
\usepackage{amsmath,amssymb,amsfonts,amscd,epsfig}
\usepackage{color}
\usepackage{graphicx}
\def\rref#1{(\ref{#1})}
\newcommand{\be}{\begin{equation}}
\newcommand{\beq}{\begin{equation}}
\newcommand{\eeq}{\end{equation}}
\newcommand{\ee}{\end{equation}}
\newcommand{\bea}{\begin{eqnarray}}
\newcommand{\eea}{\end{eqnarray}}
\newcommand{\ba}{\begin{array}}
\newcommand{\ea}{\end{array}}
\newcommand{\h}{\hat}

\newcommand{\um}{\frac{1}{2}}

\newcommand{\ga}{\gamma}

\newcommand\IZ{\mathbb{Z}}
\newcommand\IR{\mathbb{R}}

\newcommand\IT{\mathbb{T}}
\newcommand{\II}{\mathbb{I}}

\def\rref#1{(\ref{#1})}

\begin{document}

\begin{center}
{\Large\bf
Quantum Holonomies and the Heisenberg Group }\\
\vspace{.4in}
{J.\ E.~N{\sc elson}\\
       {\small\it Dipartimento di Fisica, Sezione Teorica, Universit\`a degli Studi di Torino}\\
       {\small\it and INFN Sezione di Torino}\\
       {\small\it via Pietro Giuria 1, 10125 Torino, Italy}}\\
\vspace{1ex}
{\small and}\\
\vspace{1ex}
{R.\ F.~P{\sc icken}\\
{\small\it Center for Mathematical Analysis, Geometry and Dynamical Systems}\\
{\small\it Mathematics Department}\\
{\small\it  Instituto Superior T\'{e}cnico, Universidade de Lisboa}\\
{\small\it Avenida Rovisco Pais, 1049-001 Lisboa, Portugal}}\\
\end{center}

\begin{abstract}
Quantum holonomies of closed paths on the torus $\IT^2$ are interpreted as elements of the Heisenberg group $H_1$. Group composition in $H_1$ corresponds to path concatenation and the group commutator is a deformation of the relator of the fundamental group $\pi_1$ of $\IT^2$, making explicit the signed area phases between quantum holonomies of homotopic paths. Inner automorphisms of $H_1$ adjust these signed areas, and the discrete symplectic transformations of $H_1$ generate the modular group of $\IT^2$.
\end{abstract}
\begin{flushleft}
PACS (2010) 04.60Kz - 02.20.-a\\
MSC (2010) 81R50 - 83C45
\end{flushleft}
\section{Introduction} \label{intro} \noindent 

Motivated by studies in 2+1 dimensional gravity \cite{NR1,NRZ}, in previous 
articles \cite{goldman,NP,npint,NP1} we used piecewise linear (PL) paths in 
$\IR^2$ to represent closed paths on the torus $\IT^2$. Their corresponding 
quantum holonomies were represented by quantum $SL(2,\IR)$ matrices which in general do 
not commute. The traces of holonomies corresponding to paths which intersect on 
$\IT^2$ satisfy a linear algebra \cite{goldman}, a generalization of the Goldman 
bracket \cite{gol}, where on the right hand side (RHS) there are traces of 
holonomies of paths which are rerouted in two different ways at the 
intersections. 

Rerouted paths on $\IT^2$ arising from intersections of closed paths naturally correspond to PL paths in $\IR^2$. The Heisenberg group, 
a non--commutative Lie group, has wide applications in both mathematics and physics, 
and can be interpreted geometrically in terms of PL paths on $\IR^2$ \cite{chaf}. 

Our purpose here is to present an interesting and novel relationship between quantum holonomies that represent PL paths and elements of the $3$--parameter Heisenberg group $H_1$, where $SL(2,\IR)$ matrices are replaced by elements of $H_1$. A new feature is that area phases between homotopic paths (which arise, indirectly, from a non--zero quantum curvature \cite{NP}) are now explicit. Group composition in $H_1$ and the group commutator are interpreted geometrically. Inner automorphisms of $H_1$ i.e. conjugation by a fixed element of $H_1$, adjust the signed area of a PL path (the geometric area between a PL path and a straight path between the same points) and discrete symplectic transformations generate the modular group of $\IT^2$.

The plan of the paper is as follows: in Section \ref{sec2} we review PL paths, their associated quantum holonomies, 
and the concept of signed area. In Section \ref{sec3} we present the Heisenberg group $H_1$. In Section \ref{sec4} we connect the two. 

\vskip 0.1cm

\section{PL paths and quantum holonomies}\label{sec2}
\subsection{Piecewise linear paths}\label{sub21}
Closed paths (loops) on the torus $\IT^2=\IR^2/\IZ^2$ are identified
with PL paths on its covering space ${\IR}^2$, where a PL path is between integer points $(m,n) \in \IZ^2$. All 
integer points are identified, and correspond to the same point on $\IT^2$. A path in ${\IR}^2$ representing a 
closed path on $\IT^2$ can therefore be replaced by any parallel path starting at a different integer
point. Two paths on $\IT^2$ are homotopic if and only if the corresponding paths in $\IR^2$ have the same integer
starting point and the same integer endpoint.

A natural subclass of paths in ${\IR}^2$ are those straight paths denoted
$p=(m,n)$ that start at the origin $(0,0)$ and end at an integer point
$(m,n)$. They generalize the cycles $\ga_1,\, \ga_2$
(corresponding to the paths $(1,0)$ and $(0,1)$ respectively) which satisfy
the relator of $\pi_1(\IT^2)$
\be
\ga_1^{\vphantom{-1}}\cdot \ga_2^{\vphantom{-1}}\cdot \ga_1^{-1}\cdot \ga_2^{-1}
  = {\II}.
\label{gp}
\ee

\subsection{Quantum holonomies and signed area}\label{sub22}
In previous articles \cite{goldman,NP1} a quantum $SL(2,\IR)$ matrix was assigned 
to any straight path $(m,n)$ 
\be
\hat{U}_{(m,n)}= \exp \int_{(m,n)} \hat{A}.
\label{Umn}
\ee
This extends straightforwardly to any PL path between integer points: assign a quantum matrix to each linear segment of the path, as in \rref{Umn} and multiply the matrices in the same order as the segments along the path. The general relation is
\be
p\rightarrow \hat{U}_p = {P }\exp \int_{p} \hat{A}.
\label{hol}
\ee
where $P$ denotes path-ordering. In \rref{Umn}, \rref{hol}, $\hat{A}$ is a constant connection with non-zero curvature \cite{NP}, parametrized by two non-commuting parameters ${\hat r}_1,~{\hat r}_2$. For example, for the cycles $\ga_i$ which satisfy \rref{gp}, equation \rref{hol} is
\be
\hat{U}_i = \exp \int_{\ga_i} \hat{A} = \left(\begin{array}{clcr}e^{{{\hat r}_i}}& 0 \\0& e^{-
{{\hat r}_i}}\end{array}\right) \quad i=1,2.
\label{hol2}
\ee
where $[{\hat r}_1,{\hat r}_2] = -i \hbar \frac{\sqrt {-\Lambda}}{4}$, and $\Lambda$ is a (negative) cosmological constant \cite{NR1,NRZ}.
\vskip 0.2cm

In \cite{goldman} it was shown that the holonomies \rref{Umn} satisfy the $q$--relations
\be
\h{U}_{(m,n)} \h{U}_{(s,t)} = q^{mt-ns}~ \h{U}_{(s,t)} \h{U}_{(m,n)}, \quad \h{U}_{(m,n)} \h{U}_{(s,t)} \h{U}_{(m,n)}^{-1} \h{U}_{(s,t)}^{-1} = q^{mt-ns} ~{\II}_2,
\label{uu1}
\eeq
with 
\be
q=\exp (- \frac {i \hbar \sqrt{-\Lambda}}{4}).
\label{q}
\eeq
For $\h{U}_1=\h{U}_{\ga_1}=\h{U}_{(1,0)}$, $\h{U}_2=\h{U}_{\ga_2}=\h{U}_{(0,1)}$, the $q$--relations \rref{uu1} are
\beq
 \h{U}_1  \h{U}_2 = q ~ \h{U}_2  \h{U}_1,\quad  \h{U}_1  \h{U}_2 \h{U}_1^{-1}  \h{U}_2^{-1} = q~{\II}_2
\label{fund2}
\eeq
i.e. a deformation of the holonomy relator that follows from \rref{gp}.
\vskip 0.2cm
From \rref{hol2} and the Baker-Campbell-Hausdorff identity 
\beq e^{X} e^{Y}= e^{X + Y} e^{\frac{[X, Y]}{2}}
\label{bch}
\eeq
(valid when $[X,Y]$ is a $c$--number) we also have a triangle equation
\be
\h{U}_{(m,n)} \h{U}_{(s,t)} = q^{\um(mt-ns)} \h{U}_{(m+s,n+t)}.
\label{tri}
\ee

Consider two homotopic paths on $\IT^2$ represented by two PL paths $p_1,\, p_2$ on 
$\IR^2$. It was shown in \cite{goldman} that the following relationship holds for the respective quantum matrices:
\begin{equation}
\hat{U}_{p_1}=q^{S(p_1,p_2)}\hat{U}_{p_2},
\label{areaphase}
\end{equation}
where $S(p_1,p_2)$ denotes the signed area enclosed between the paths
$p_1$ and $p_2$, and $q$ is the quantum phase \rref{q}. The signed area between two
PL paths is defined as follows: for any finite region $R$ enclosed by $p_1$ and
$p_2$, if the boundary of $R$ consists of oriented segments of $p_1$ and
$p_2^{-1}$ (the path $p_2$ followed in the opposite direction), and is globally
oriented in the positive (anticlockwise), or negative (clockwise) sense, this
gives a contribution of $+{\rm area}(R)$ , or $-{\rm area}(R)$ respectively, to
the signed area $S(p_1,p_2)$ (otherwise the contribution is zero).

\section{The Heisenberg group $H_1$}\label{sec3}
The Heisenberg algebra $h_1$ is a 3--dimensional Lie algebra with generators $X, Y, Z$ satisfying the commutation relations
\beq 
[X,Y] = XY-YX = Z , \quad [X,Z] = [Y,Z] = 0.
\label{hcomm}
\eeq
The $3$-parameter Heisenberg group $H_1$ has elements 
\beq
(a,b,c) = \exp {(aX + bY + cZ)}
\label{helem}
\eeq
where $a,b,c$ are canonical, or exponential coordinates, and $X,Y,Z$ satisfy \rref{hcomm}. If $a,b,c~\in~\IZ$ then $H_1$ is
the discrete Heisenberg group. Otherwise it is the continuous Heisenberg group. The third element $c$ in \rref{helem} is a {\it phase}, and 
from \rref{hcomm} it commutes with all elements of $H_1$.
\vskip 0.2cm
\noindent We list some selected properties \cite{kis}:
\begin{enumerate}

\item The identity in $H_1$ is $(0,0,0)$ and the inverse of $(a,b,c)$ is $(-a,-b,-c)$. 
\item The group composition is, from \rref{bch}, \rref{hcomm} and \rref{helem}
\beq
(a,b,c)*(a^{\prime}, b^{\prime},c^{\prime}) = (a+a^{\prime},b+b^{\prime},c+c^{\prime}+\um (a b^{\prime}-a^{\prime} b)).
\label{hcomp}
\eeq
\item The center of $H_1$ consists of elements $\{(0,0,c)\}$, $c~\in~ \IR$. An example is the group commutator 
\beq 
(a,b,c)*(a^{\prime}, b^{\prime},c^{\prime})*(a,b,c)^{-1} * (a^{\prime}, b^{\prime},c^{\prime})^{-1} 
\label{gcomm}
\eeq
which from \rref{hcomp} is equal to $(0,0,a b^{\prime}-a^{\prime} b)$.

\end{enumerate}

\subsection{Automorphisms of $H_1$}
The map $A: ~f \rightarrow A(f)$, with $f,~ A(f)~\in~ H_1$ is an automorphism of $H_1$ if $A(f*g)=A(f)*A(g)$, where $*$ is group composition. The automorphisms of $H_1$ are \cite{kis}
\begin{itemize}
 \item Inner i.e. conjugation by a fixed element of $H_1$. This adds a phase since from \rref{hcomp}
\beq
(x^{\prime}, y^{\prime}, z^{\prime}) \rightarrow (x, y, z)*(x^{\prime}, y^{\prime}, z^{\prime})*(x, y, z))^{-1}
 =(x^{\prime},y^{\prime},z^{\prime}+xy^{\prime}-yx^{\prime}).
\label{autin}
\eeq
\item Outer i.e.

Dilations $(x,y,z) \rightarrow (rx,ry,r^2 z),\quad r>0,\quad r~\in~\IR$.

Inversions $(x,y,z) \rightarrow (y,x,-z)$ (a reflection about the $x-y$ diagonal).

Symplectic 
\beq 
\left( \begin{array}{c}
x\\
y 
\end{array} \right)
\rightarrow A \left( \begin{array}{c}
x\\
y 
\end{array} \right) \quad A ~\in ~SL(2,\IR)
\label{symp}
\eeq
and do not affect $z$ i.e. there is no change of phase. Special cases are the modular transformations, $A ~\in ~SL(2,\IZ)$, whose generators $S$ and $T$ are
\beq
S=\left( \begin{array}{cc}
0&-1\\
1&0 
\end{array} \right),
\quad T=\left( \begin{array}{cc}
1&0\\
1&1 
\end{array} \right),
\label{mod1}
\eeq
and satisfy $S^2=-\II,(ST)^3=\II$.

Other symplectic transformations are rotations by an angle $\theta$ in the $x-y$ plane
\beq
A=\left( \begin{array}{cc}
\cos \theta & -\sin \theta\\
\sin \theta & \cos \theta 
\end{array} \right).
\label{rot}
\eeq 
\end{itemize}

\section{Quantum holonomies as elements of $H_1$}\label{sec4}
\begin{figure}[hbtp]
\centering
\includegraphics[height=1.5cm]{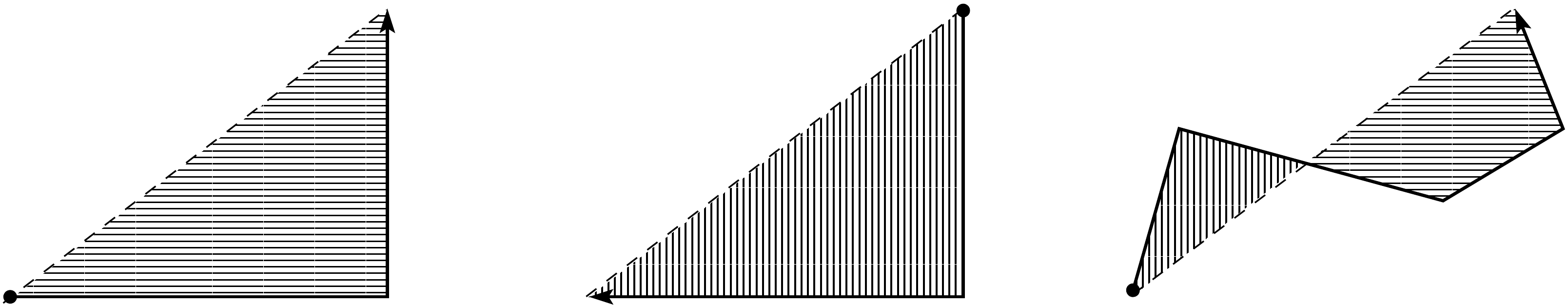}
\caption{A PL path, its inverse, a general PL path, their chords (the dashed lines) and signed areas. Horizontal shading denotes positive areas and vertical shading denotes negative areas.}
\label{crooked}
\end{figure}

Our notation is as follows: holonomies represented by quantum $SL(2,\IR)$ matrices are denoted $\h{U}$. Elements of $H_1$ are $(a,b,c)$ as in \rref{helem}. Holonomies, denoted $U$ or $U_p$, are assignments of an element of $H_1$ to a PL path $p$. One advantage of this assignment is that the phase $c$ is now explicit {\it in the elements of $H_1$}, and geometrically is equal to the signed area in $\IR^2$ between the path $p$ and its chord (the straight line connecting the integer points $(0,0)$ and $(a,b)$). See Figure \ref{crooked}. Group composition and the group commutator have natural geometrical interpretations. This construction is related to \cite{bal} where it was shown that, for a curve $\ga$ from the origin to $(a,b)$ in $\IR^2$, there is a unique lifted curve in $H_1$ which connects the identity $(0,0,0)$ to an element $(a,b,c)$, where $c$ is the signed area between $\ga$ and its chord. 
\vskip 0.2cm
\noindent Note that for PL paths we require that $a,b ~\in~ \IZ$, but not necessarily $c$. 

\noindent 
Quantum holonomies \rref{hol} are represented by elements of $H_1$, with properties 
\begin{enumerate}

\item The identity in $H_1$ $(0,0,0)$ is the identity holonomy (representing the point $(0,0)$, or the trivial path with zero signed area). The inverse of an element in $H_1$ is the inverse holonomy (and corresponds to the inverted PL path). See Figure \ref{crooked}. 
 
\item Group composition \rref{hcomp} corresponds to path concatenation. To see this, consider the composition of two elements $(a,b,c)$ and $(a^{\prime},b^{\prime},c^{\prime})$ in $H_1$. 

This concatenation is shown in the first two figures of Figure \ref{triangle}. The signed area between the resulting PL path and its chord $(a+ a^{\prime},b + b^{\prime})$ is $c+c^{\prime}+\um (a b^{\prime}-a^{\prime} b)$. It is easily seen that $\um (a b^{\prime}-a^{\prime} b)$ is the signed area between the two concatenated straight paths ($(a,b)$, and the straight path from the point $(a,b)$ to the point $(a + a^{\prime},b + b^{\prime})$), and their chord (also $(a+ a^{\prime},b + b^{\prime})$), shown on the right of Figure \ref{triangle}.

For straight paths ($c=c^{\prime}=0$) group composition \rref{hcomp} corresponds to the triangle relation \rref{tri}.

\begin{figure}[hbtp]
\centering
\includegraphics[height=2cm]{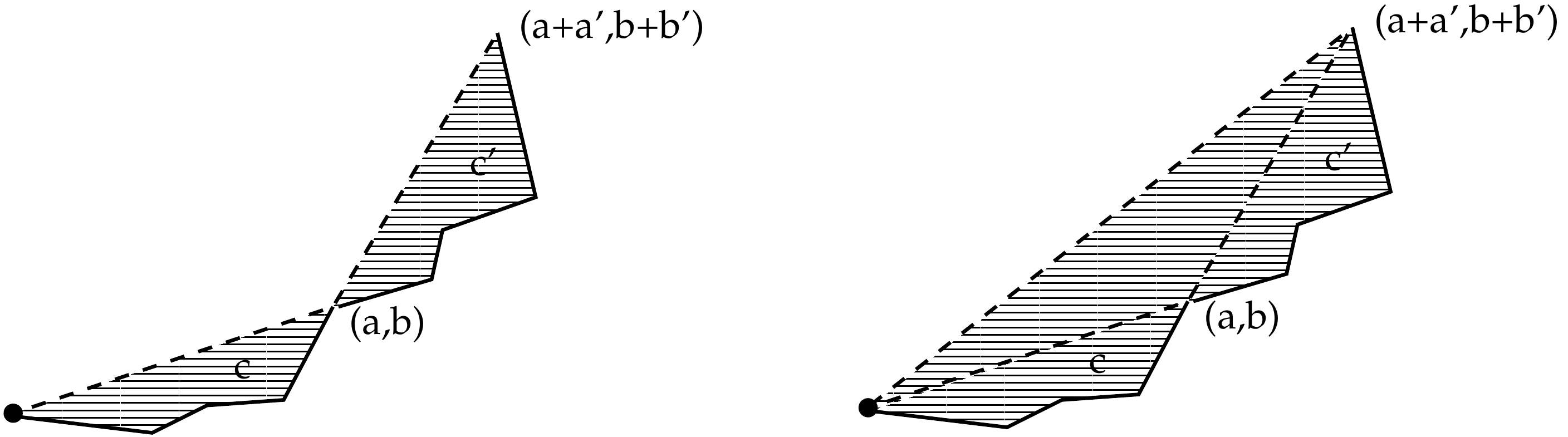}
\includegraphics[height=2cm]{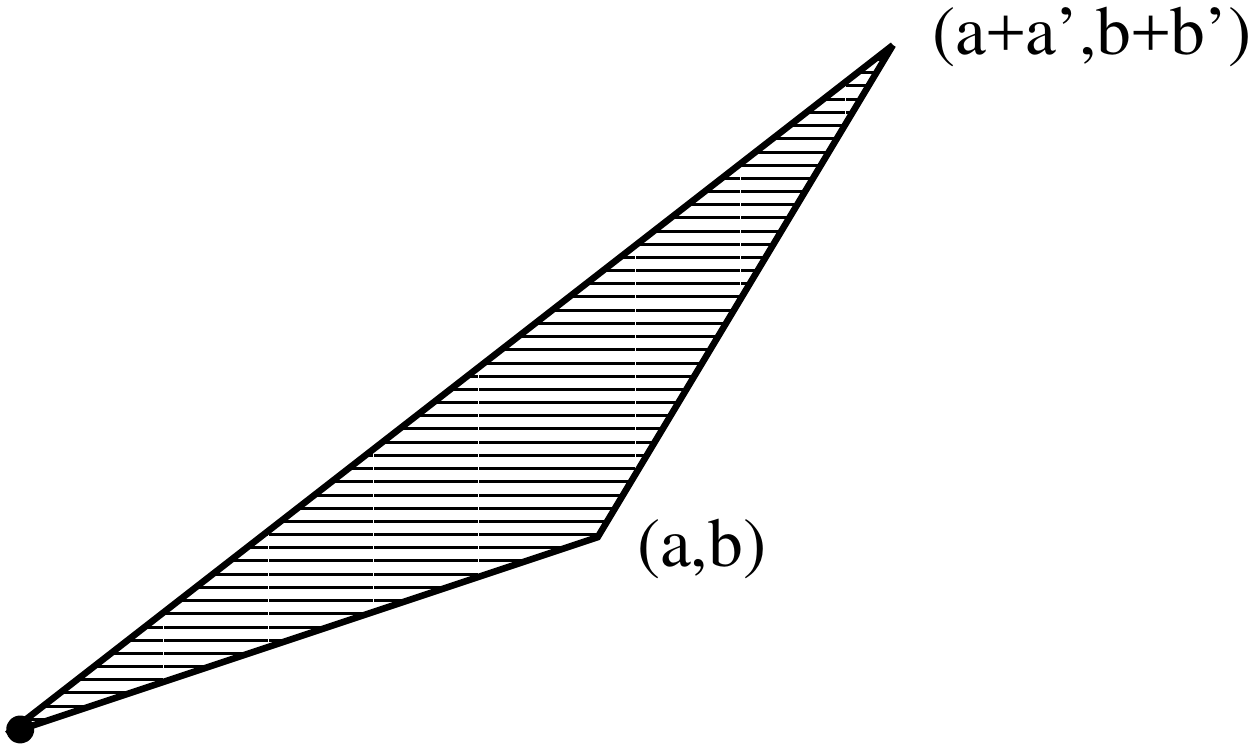}
\caption{Concatenation of two PL paths and two straight paths.}
\label{triangle}
\end{figure}

\item Homotopy and area phases. Consider two elements $U_{p_1} = (x,y,z)$ and $U_{p_2} = (x,y,z^{\prime})$ of $H_1$. The group 
composition \rref{hcomp} of $U_{p_1}$ and ${U_{p_2}}^{-1}$  is
\beq
(x,y,z) * (-x,-y,-z^{\prime}) = (0,0,z-z^{\prime}) = e^{(z-z^{\prime})Z}~(0,0,0).
\label{ppinv}
\eeq
The RHS of equation \rref{ppinv} can be understood in terms of the signed area enclosed between two PL paths  
$p_1$ and $p_2$ with the same starting point $(0,0)$ and endpoint $(x,y)$ but with different phases.  In terms of holonomies 
$$U_{p_1} =(x,y,z)=e^{zZ}~U_{(x,y)}, ~ U_{{p_2}^{-1}}= (-x,-y,-z^{\prime})=e^{-z^{\prime}Z}~U_{(-x,-y)}$$ 
and using the triangle relation \rref{tri}
\beq 
U_{p_1} ~U_{{p_2}^{-1}} =e^{(z-z^{\prime})Z} ~ U_{(0,0)}= (0,0,z-z^{\prime})
\label{pl2}
\eeq
in agreement with \rref{ppinv}. Further, from \rref{pl2} we deduce that 

\noindent $U_{p_1} =e^{(z-z^{\prime})Z}~U_{p_2}$, i.e. the PL paths $p_1$ and $p_2$ are homotopic, with relative signed area $z-z^{\prime}$.

An element $(a,b,c)$ of $H_1$ corresponds to an equivalence class of homotopic paths which have zero relative area phase. Consider e.g. the 
chord $(a,b,0)$. Any homotopic zig-zag path with equal area above and below the chord has zero area phase with the chord. Therefore all such paths 
also have zero relative area phase, since signed area is additive \cite{NP} for composition of homotopies of paths.

\item The center of $H_1$ consists of elements $\{(0,0,c)\}$, $c~\in~ \IR$, interpreted as the holonomies $U_p$ of closed paths $p$ in 
$\IR^2$, where $c$ is the enclosed area. An example is the group commutator \rref{gcomm} which corresponds to the closed path 
$p=p_1p_2{p_1}^{-1}{p_2}^{-1}$, with $U_{p_1}=(a,b,c)$, $U_{p_2}= (a^{\prime},b^{\prime},c^{\prime})$ shown in Figure \ref{centre}. The 
enclosed area is $a b^{\prime}-a^{\prime} b$. 
 
\begin{figure}[hbtp]
\centering
\includegraphics[height=1.5cm]{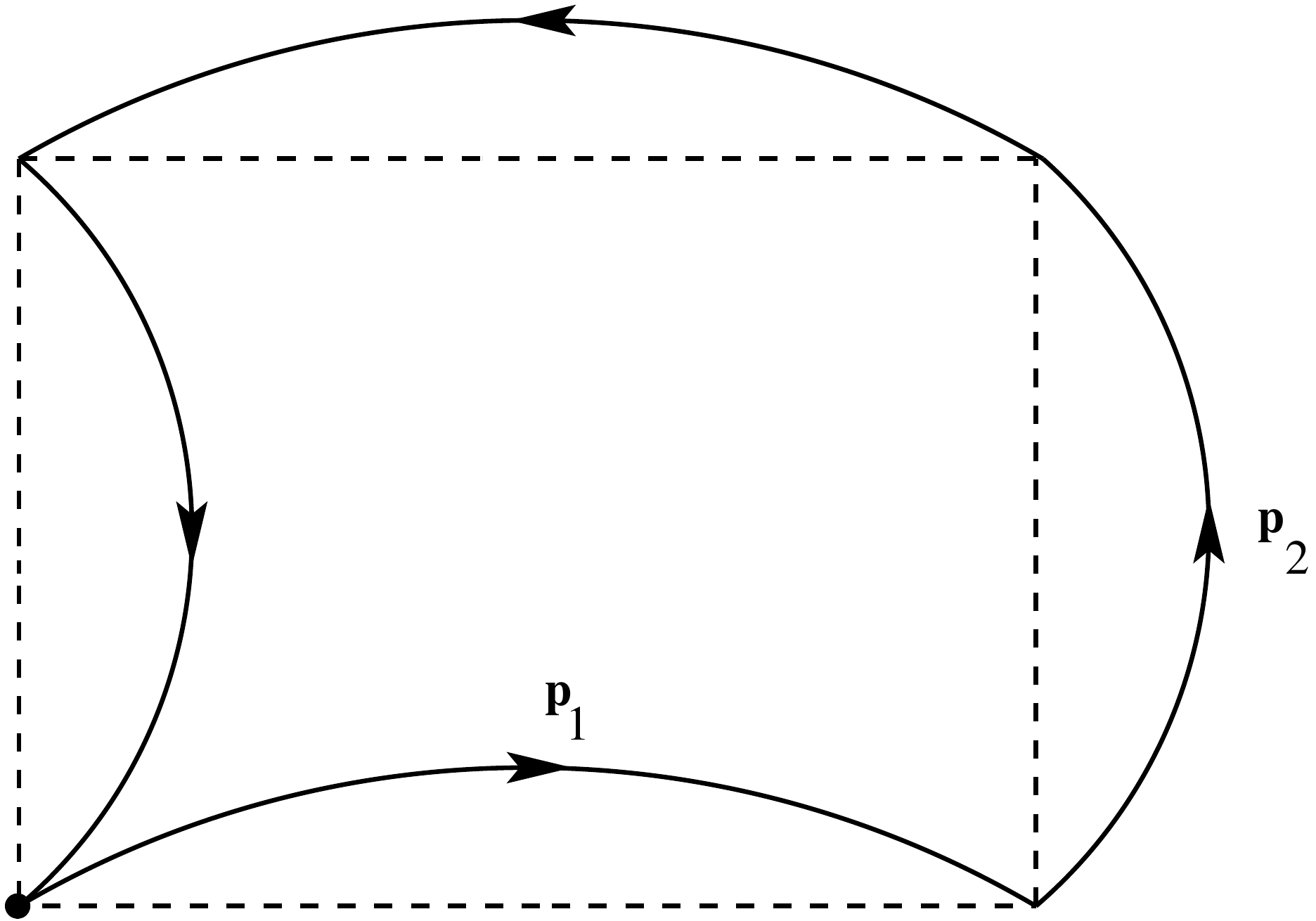}
\caption{The closed path $p_1p_2{p_1}^{-1}{p_2}^{-1}$. }
\label{centre}
\end{figure}

The group commutator \rref{gcomm} also describes the deformed relator \rref{fund2}. 

\item 
It is natural to interpret signed area phases in terms of crossed modules and surface transport in higher gauge theory \cite{hgt}. There 
is a crossed module given by the inclusion homomorphism $[H_1,H_1] \rightarrow H_1$, with trivial action of $H_1$ on $[H_1,H_1]$, where 
$[H_1,H_1]$ denotes the commutator subgroup of $H_1$ (the subgroup generated by commutators). Here it is the subgroup of elements
$\{(0,0,c)\}$, $c~\in~ \IR$, i.e. the center of $H_1$.

\item Automorphisms.
\begin{itemize}

 \item Inner automorphisms - from \rref{autin}, conjugation by an element of $H_1$ adds a phase. Geometrically, this can be interpreted as the 
 action of a path $p_1$ on a path $p_2$, i.e. $p_2 \mapsto  p_1 p_ 2 p_1^{-1}$ which adjusts the signed area between $p_2$ and its chord, as can 
 be seen from Figure \ref{centre}.
 
\item Outer automorphisms - the symplectic transformations in $H_1$ with $A ~\in ~SL(2,\IZ)$ \rref{mod1} generate the modular group (with $q=e^Z$)
\begin{align}
S:&~U_1 \rightarrow U_2, \quad  U_2  \rightarrow {U_1}^{-1}\nonumber\\
T:&~U_1 \rightarrow q^{-\um}~U_1 ~U_2, \quad U_2 \rightarrow U_2,
\label{modhol}
\end{align}
a symmetry group of the deformed relator \rref{fund2}.
\end{itemize}
\end{enumerate}

\section*{Final Remarks}
We have found a novel relationship between quantum holonomies representing PL paths and elements of the Heisenberg 
group $H_1$. Group composition corresponds to path concatenation, the group commutator describes a deformation of 
the relator of the fundamental group $\pi_1$ of $\IT^2$, and there are interpretations in terms of homotopy and 
area phases. Inner automorphisms of $H_1$ correspond to adjusting the signed area of a PL path, and discrete 
symplectic transformations generate the modular group.

The triangle relation \rref{tri} is reminiscent of a quantum bracket for traces of holonomies of straight paths 
derived in \cite{goldman} 
\begin{align}
&[ \h{T}(m,n), \h{T}(s,t)] = \h{T}(m,n) \h{T}(s,t) - \h{T}(s,t) \h{T}(m,n)\nonumber\\
& = (q^{\um (mt-ns)} - q^{-\um (mt-ns)})~(\h{T}(m+s,n+t) - \h{T}(m-s,n-t)),
\label{tcomm} 
\end{align}
where $\h{T}(m,n)$ is the trace of the holonomy $\h{U}_{(m,n)}$. If $\h{U}_{(m,n)} \in SL(2,\IR)$ this is 
\beq
\h{U}_{(m,n)}+\h{U}_{(m,n)}^{-1}=\h{U}_{(m,n)}+ \h{U}_{(-m,-n)}= tr ~\h{U}_{(m,n)}~\II_2.
\label{tstr}
\eeq
The RHS of \rref{tcomm} contains two terms, corresponding to paths rerouted in two different ways at the intersections of 
the paths represented by $\h{U}_{(m,n)}$ and $\h{U}_{(s,t)}$. However, the RHS of \rref{tri} contains only one rerouting. To 
make contact with the bracket \rref{tcomm}, it would be necessary to analyse the intersections of arbitrary PL paths, and 
to include inverses of holonomies, as for example, in \rref{tstr}. 

Holonomies in 2+1 gravity \cite{goldman,NP} were represented by quantum $SL(2,\IR)$ matrices (with q--number entries, composed by matrix multiplication). Representations of $H_1$ are by $3 \times 3$ matrices (with $c$-number entries and matrix multiplication) or by differential operators (with operator composition). The correlation presented here is, however, independent of the representations of either. 

The above issues, their applications and implications will be discussed elsewhere \cite{NPnext}.

\section*{Acknowledgements}

This work was supported by INFN (Istituto Nazionale di Fisica Nucleare) Iniziativa Specifica GSS, 
MIUR (Ministero dell' Universit\`a e della Ricerca Scientifica e Tecnologica) (Italy) and FCT 
(Funda\c{c}\~{a}o para a Ci\^{e}ncia e a Tecnologia) project UID/MAT/04459/2013 (Portugal).

\end{document}